\documentclass[12pt]{article}
\usepackage{amsmath}
\usepackage{amssymb}
\usepackage{latexsym}
\usepackage{epsfig}
\usepackage{graphicx}
\usepackage{epstopdf}
\usepackage{amssymb}
\usepackage{euscript}

\catcode`\@=11 \@addtoreset{equation}{section}

\def\e{{\epsilon}}

\def\p{{\bf p}}

\def\bra#1{\langle #1 |}
\def\ket#1{|#1 \rangle}
\def\0{\nonumber}

\def\cos{{\rm cos}}

\def\tr{{\rm tr}}
\def\Tr{{\rm Tr}}

\def\ch{{\cal{H}}}

\newcommand\ee{\end{eqnarray}} 
\newcommand\be{\begin{eqnarray}}
\newcommand\ba{\begin{array}} 
\newcommand\ea{\end{array}}
\newcommand\eeq{\end{equation}} 
\newcommand\beq{\begin{equation}}

\def\Tr{ \hbox{\rm Tr}}

\def\bra{\langle}\def\ket{\rangle}

\def\dag{{}^{\dagger}}

\def\p{{}^{\prime}}

\def\hsp{,\hspace{.7cm}}
\newcommand{\C}{\ensuremath{\mathbb C}}
\newcommand{\Z}{\ensuremath{\mathbb Z}}
\newcommand{\R}{\ensuremath{\mathbb R}}

\newcommand{\oC}{\ensuremath{\overline{C}}}
\newcommand{\oA}{\ensuremath{\overline{A}}}
\newcommand{\metastable}{unstable}

\begin{document}
\begin{flushright}
ULB-TH/07-02\\ hep-th/0702042
\end{flushright}

\vspace{.1in}
\begin{center}
{\Large\bf Puffed Noncommutative Nonabelian Vortices} \\
\vspace{5mm}
\end{center}
\vspace{0.1in}
\begin{center}
Nazim Bouatta $^{(a)}$\footnote{Chercheur FRIA,
nbouatta@ulb.ac.be}, Jarah Evslin $^{(a)}$\footnote{
jevslin@ulb.ac.be}, Carlo Maccaferri
$^{(b)}$\footnote{carlo.maccaferri@ulb.ac.be} \vspace{7mm}

$^{(a)}$ {\it Physique Th\'eorique et Math\'ematique, Universit\'e
Libre de Bruxelles \& International Solvay Institutes, ULB Campus
Plaine C.P. 231, B--1050 Bruxelles, Belgium}\vspace{5mm}

$^{(b)}$ {\it Theoretische Natuurkunde, Vrije Universiteit
Brussel, Physique Th\'eorique et Math\'ematique, Universit\'e
Libre de Bruxelles, and The International Solvay Institutes
Pleinlaan 2, B-1050 Brussels, Belgium }
\end{center}
\vspace{0.6in}
\begin{center}
{\bf Abstract}
\end{center}

\noindent We present new solutions of noncommutative gauge
theories in which coincident \metastable\ vortices expand into
\metastable\ circular shells. As the theories are noncommutative,
the naive definition of the locations of the vortices and shells
is gauge-dependent, and so we define and calculate the profiles of
these solutions using the gauge-invariant noncommutative Wilson
lines introduced by Gross and Nekrasov. We find that charge 2
vortex solutions are characterized by two positions and a single
nonnegative real number, which we demonstrate is the radius of the
shell. We find that the radius is identically zero in all
2-dimensional solutions. If one considers solutions that depend on
an additional commutative direction, then there are time-dependent
solutions in which the radius oscillates, resembling a braneworld
description of a cyclic universe. There are also smooth BIon-like
space-dependent solutions in which the shell expands to infinity,
describing a vortex ending on a domain wall.


\newpage
\section{Introduction}
Noncommutative gauge theories with or without adjoint scalars
and/or fundamental fermions are known to admit \metastable\ vortex
solutions. The shapes and positions of these noncommutative
vortices are difficult to describe or even define, as the
positions of the gauge field, scalars and fermions are gauge
dependent. In Ref.~\cite{GN} the authors introduced two kinds of
gauge-invariant operators, a fermion bilinear and also a trace of
the gauge field in momentum space which allow one to define and
measure the positions of these configurations in a gauge-invariant
way. They arrived at the interesting conclusion that the positions
of a set of vortices are the eigenvalues of its Weyl-transformed
Wilson lines.

In this note we demonstrate that, if a vortex charge is greater
than one, then there are new gauge-invariant quantities which are
distinguished by both gauge-invariant operators. We find that the
equations of motion demand that these quantities vanish in the
2-dimensional theories described in Ref.~\cite{GN}, but not in
higher-dimensional analogues. We find the positions of these new
solutions by calculating the Fourier transform of the trace of the
momentum space wavefunction, which following \cite{GN} is defined
to be a kind of noncommutative Wilson loop.

In the case of coincident charge 2 vortices, we find solutions in
which a codimension 2 vortex expands into a thin circular domain
wall reminiscent of the commutative solitons in
Refs.~\cite{Stefano,Stefano2,Stefano3}. Unlike non-BPS semilocal
vortices \cite{semilocal,semilocal2} in commutative gauge
theories, the domain wall appears to have a sharp outer boundary,
although a change in normalization of the Wilson loops would
smoothen this boundary. There is precisely one new nonnegative
real gauge-invariant quantity in the charge 2 case, which we show
corresponds to the radius of this shell. The radius is
proportional to the commutator of the gauge fields in the
transverse directions, analogously to the construction of higher
dimensional D-branes from lower dimensional D-branes with
noncommuting position matrices, in keeping with the identification
of the gauge field eigenvalues and the positions in
Refs.~\cite{AGMS,GN}. In solutions with dependence on an
additional commutative spatial dimension, this circle grows to
infinity, reaching infinity at a finite position, and so describes
a vortex ending on a domain wall. If instead one considers
time-dependent solutions with 2 noncommutative spatial dimensions,
then the radius oscillates periodically in the commutative time
direction. If one includes fundamental fermions, then when the
radius vanishes there are two fermion zero modes, of which one is
lifted by a nonvanishing radius.

\section{Vortices in the 2d theory}

\subsection{The gauge theory}

Consider bosonic string theory on the space $M^{24}\times\R^2$
where $M^{24}$ is an arbitrary 24-manifold. Wrap a single
spacefilling D25-brane on the entirety of spacetime, and consider
a $B$-field that is constant and has both legs along $\R^2$. The
spectrum of open strings ending on this D25-brane includes
massless vectors which are gauge bosons for a $U(1)$ gauge
symmetry.

The $U(1)$ gauge bundle has a curvature $F$. However the open
strings couple to $F$ in the combination $B+F$, which means that
covariant derivatives acting on charged fields have contributions
not only from the vector potential $iA$ of $F$, but also for that
of $B$. We will refer to the total vector potential as $C$. In
particular, even if $F=0$ then the wavefunction of a particle
traveling around a loop $\gamma$ gains a phase equal to the
loop-integral of $C$ which is equal by Stoke's theorem to the
integral of $B$ on the interior $D$ of the loop \beq \int_\gamma
C=\int_D dC=\int_D B+F=\int_D B \eeq where $\partial D=\gamma$.
Equivalently, if $x$ and $y$ are a set of coordinates on $\R^2$,
then if a charged particle moves first in the $x$ direction and
then in the $y$ direction, its phase will not be the same as if it
moved first in the $y$ direction and then in the $x$ direction.
For this reason this theory is called a noncommutative $U(1)$
gauge theory. In particular, the fact that translations in $x$ and
$y$ do not commute means that the $x$ and $y$ translation
generators, the momenta, do not commute. In turn this implies that
the operators $x$ and $y$ themselves do not commute.

Although the $\R^2$ direction is noncompact, if we are interested
in solutions which are normalizable in the $\R^2$ direction we may
dimensionally reduce the theory to a 24-dimensional theory on
$M^{24}$. Consider two of the 26-dimensional fields, the adjoint
scalar $\phi$ and the $U(1)$ gauge field $A$. At each point in the
24-dimensional space $M^{24}$, $\phi$ and $A$ are functions of $x$
and $y$. As they are normalizable, these functions may be expanded
in terms of Laguerre polynomials, which are a countably infinite
basis of functions. The coefficients of these polynomials may be
arranged in two infinite-dimensional matrices, which by an abuse
of notation we will also refer to as $\phi$ and $A$ such that the
noncommutativity of the $x$ and $y$ dependence is captured by the
noncommutativity of matrix multiplication. We can even write the
full connection $C$ as a matrix. These infinite-dimensional
matrices are known as the Weyl transforms of the $\R^2$-dependent
fields.

$iA$ and $C$ are both vectors, and so are described by 26
Hermitian matrices, one for each component. We will define \beq
A=A_x+iA_y\hsp \overline{A}=A_x-iA_y\hsp C=C_x+iC_y\hsp\
\overline{C}=C_x-iC_y. \eeq Rescaling the $\R^2$ coordinates so
that the commutator of $x$ and $y$ is equal to $i$, corresponding
to $\theta=1$ in the usual parametrization, one finds that $x+iy$
and $x-iy$ represent the generators $a$ and $a^\dagger$ of the
Heisenberg algebra. Therefore we may choose a convention in which
\beq C=a\dag-iA\hsp \overline{C}=a+i\overline{A}. \eeq Recall that
$iA_x$, $iA_y$, $C_x$ and $C_y$ are Hermitian, and so $iA$,
$i\overline{A}$, $C$ and $\overline{C}$ will generally not be
Hermitian and need not even be diagonalizable.

The Weyl transformed $A$ is the connection for a 24-dimensional
$U(\ch)$ gauge theory. If all fields, like $A$ and $\phi$,
transform in the adjoint of the gauge group then the center
$U(1)\subset U(\ch)$ acts trivially. This means that the fields
form representations of the smaller gauge group \beq
PU(\ch)=\frac{U(\ch)}{U(1)} \eeq and only the $PU(\ch)$ gauge
bundle needs to be well-defined. The effective gauge group
$PU(\ch)$ is an Eilenberg-MacLane space with nontrivial homotopy
group \beq \pi_2(PU(\ch))=\Z \eeq and so nontrivial gauge bundles
are characterized entirely by an integral 3-class $H$, which will
be identified with the NS 3-form.

The gauge groups $U(\ch)$ and $PU(\ch)$ only appear after the
dimensional reduction from 26 to 24 dimensions, and so it appears
that the gauge bundle is fibered over only a 24-dimensional
subspace of the 26-dimensional spacetime, although in principal
its characteristic class $H$ is defined on the entire bulk. This
is because we chose to start with a single D25-brane. The Sen
conjecture \cite{Sen} has taught us that the open strings on a
single D25-brane do not capture all of the physics, one needs an
infinite stack. In the AdS/CFT correspondence
\cite{Mald,GKP,Witads} this corresponds to the fact that the open
strings that end on an infinite stack of D-branes know everything
about the closed string sector. Therefore to capture all of the
information about the string theory, one would have needed an
infinite stack of D25-branes, which would have led to the desired
$PU(\ch)$ bundle over the bulk. This may appear to be in
contradiction with the possibility that the $H$ flux is nontrivial
around some cycle, which would disallow a spacefilling brane,
indeed the Sen picture breaks down in that case and it is harder
to find the closed strings in the open string physics.


We will first consider configurations which are constant on
$M^{24}$, yielding a 2-dimensional Euclidean gauge theory which is
dimensionally reduced to a 0-dimensional matrix model via the Weyl
transform. Let \beq w=\frac1{\sqrt2}(x+iy)\hsp
\overline{w}=\frac1{\sqrt2}(x-iy) \eeq be complex coordinates on
the $\R^2$. We will use the symbols $\partial$ and
$\overline{\partial}$ for derivatives which are covariant with
respect to the connection of the $B$ field but not the gauge field
in the directions $w$ and $\overline{w}$ respectively. We have
seen that these represent the usual raising and lower operators in
the Heisenberg algebra.

Using these derivatives and the $U(1)$ gauge potential $A$, which
is dimensionally reduced to a $U(\ch)$ gauge potential in the
matrix model, we may define a $U(\ch)$ field strength \beq
F_{w\overline{w}}=\partial\overline{A}-\overline{\partial}A-i[A,\overline{A}]=-i[C,\overline{C}]-i.
\label{mag} \eeq In terms of this field strength, and the adjoint
scalar $\phi$ we may write the action for our matrix model
following, for example, Ref.~\cite{Harvey} \beq
S=2\pi\Tr_{\ch}(-\frac{1}{4}F_{ij}F^{ij}-[C,\phi][\overline{C},\phi]-V(\phi))
\eeq where $V(\phi)$ is a potential function with a local maximum
at $\phi=0$ and a local minimum at $\phi=1$.

We may now obtain equations of motion by varying the action with
respect to $\phi$, $C$ and $\overline{C}$. In the square of $F$,
the $[C,\overline C]$ and $1$ terms are both topological and so do
not contribute to the equations of motion, thus we need only
consider the $[\overline{C},C]^2$ term.

Varying $\phi$ one obtains the equation of motion \beq \label{feq}
0=[\overline{C},[C,\phi]]-V\p(\phi). \eeq
Varying $C$ one finds \beq \label{ceq}
[\overline{C},[\overline{C},C]-[\phi,[\overline{C},\phi]]=0 \eeq
and varying $\overline{C}$ we find its transpose \beq \label{cbeq}
[C,[C,\overline{C}]-[\phi,[C,\phi]]=0. \eeq Now we are ready to
choose an ansatz and solve these equations. The adjoint scalar
will not play a crucial role in the solutions that we will
present, they will all have analogues in a truncated theory in
which one omits the $\phi$ field entirely.

\subsection{Two-dimensional solutions and symmetries}

We will be interested in an solutions describing $N$ point-like
branes. This corresponds to the ansatz \beq \phi=\phi_*(1-P_N)
\eeq where $1$ is the identity matrix, $P_N$ is the projector onto
an $N$-dimensional subspace $\C^N\subset\ch$ and $\phi_*$ is the
stable minimum of $V(\phi)$. In Ref.~\cite{GMS} the authors
demonstrated that in this ansatz the potential term vanishes in
the $\phi$ equation of motion (\ref{feq}).

The projector decomposes the Hilbert space $\ch$ into its
eigenspaces, a $\C^N$ which it annihilates and the remaining $\ch$
on which it has eigenvalue one. We can use this decomposition to
decompose $C$ and $\overline{C}$ in terms of an $N\times N$, an
$N\times\infty$, an $\infty\times N$ and an $\infty\times\infty$
submatrix \beq C = \left(
\begin{array}{cc}
\alpha & \beta\\
\gamma & \delta\\
\end{array}\right)\hsp
\overline{C} = \left(
\begin{array}{cc}
\alpha^\dagger & \gamma^\dagger\\
\beta^\dagger & \delta^\dagger\\
\end{array}\right).
\eeq

Now we will insert this decomposition into the equations of
motion. In terms of the decomposition we can evaluate the
commutators \beq [C,\phi]=\left(
\begin{array}{cc}
0 & \beta\\
-\gamma & 0\\
\end{array}\right)\hsp
[\overline{C},\phi] = \left(
\begin{array}{cc}
0 & \gamma^\dagger\\
-\beta^\dagger & 0\\
\end{array}\right).
\eeq Inserting these commutators into the $\phi$ equation of
motion (\ref{feq}) we find \beq \label{commi}
0=[\overline{C},[C,\phi]]=\left(
\begin{array}{cc}
-\gamma^\dagger\gamma-\beta^\dagger\beta & \alpha\dag\beta-\beta\delta\dag\\
-\delta\dag\gamma+\gamma\alpha\dag & \beta\dag\beta+\gamma\dag\gamma\\
\end{array}\right).
\eeq The upper left entry is negative-definite and the lower right
entry is positive-definite. Neither is zero unless every component
of the matrices $\beta$ and $\gamma$ vanishes, which leaves \beq
C= \left(
\begin{array}{cc}
\alpha & 0\\
0 & \delta\\
\end{array}\right)\hsp
\overline{C} = \left(
\begin{array}{cc}
\alpha^\dagger & 0\\
0 & \delta^\dagger\\
\end{array}\right).
\eeq When $\beta=\gamma=0$ every block in (\ref{commi}) vanishes
and so Eq.~(\ref{feq}) is satisfied.

Next we need to solve the $C$ and $\overline{C}$ equations of
motion (\ref{ceq}) and (\ref{cbeq}). The fact that $C$ and
$\overline{C}$ are block-diagonal means that these equations of
motion can be decomposed into the equations of motion for the
$N\times N$ block and the equations of motion for the
$\infty\times\infty$ block, which each need to be solved
separately.

We will start with the easier, finite-dimensional $N\times N$
block. The $\overline{C}$ equation of motion (\ref{cbeq}) for this
block is \beq 0=[\alpha,[\alpha\dag,\alpha]]. \label{commutano}
\eeq Note that $[\alpha\dag,\alpha]$ is diagonalizable because it
is Hermitian. Choose a basis for $\C^N$ in which
$[\alpha\dag,\alpha]$ is diagonal. Now divide $\C^N$ into two yet
smaller spaces $\C^J$ and $\C^K$ such that $\C^J$ is the zero
eigenspace of $[\alpha\dag,\alpha]$. We can rescale the
coordinates in $\C^K$ so that $[\alpha\dag,\alpha]$ is the
$K\times K$ identity matrix. $\alpha$ also respects this block
diagonalization as a result of Eq.~(\ref{commutano}), and
therefore so does $\alpha\dag$. This means that $\alpha$ and
$\alpha\dag$ generate a $K$-dimensional representation of the
Heisenberg algebra. The Heisenberg algebra only has
representations in dimension $0$ and $\infty$. If we assume that
$N$ is finite, so that we are looking for stacks of finite numbers
of solitons, then $K\leq N$ and so $K$ is also finite. Therefore
$\alpha$ and $\alpha\dag$ generate the zero-dimensional
representation of the Heisenberg algebra, so $K=0$ and $J=N-K=N$.
This means that the zero eigenspace of $[\alpha\dag,\alpha]$ is
all of $\C^N$, and so \beq [\alpha\dag,\alpha]=0 \eeq in other
words $\alpha$ and $\alpha\dag$ are simultaneously diagonalizable
\cite{DN}. We will name their eigenvalues $\alpha_i$ and
$\overline{\alpha}_i$ respectively. When we consider solutions
with dependence on commutative directions, Eq.~(\ref{commutano})
will no longer be satisfied and we will find that $\alpha$ and
$\alpha\dag$ do not necessarily commute.

Next we treat the lower-right $\infty\times\infty$ block. Now the
$\overline{C}$ equation of motion (\ref{cbeq}) is \beq
0=[\delta,[\delta\dag,\delta]] \label{ddd} \eeq which implies that
$\delta$ and $[\delta\dag,\delta]$ are simultaneously
diagonalizable. While they can be simultaneously diagonalized, in
what we will identify as the coherent state basis, we will not
diagonalize them. Instead, we recall that \beq C=a\dag-iA\hsp
\overline{C}=a+i\overline{A} \label{Cdef} \eeq and so if we are
interested in configurations in which the $U(1)$ gauge fields $A$
and $\overline{A}$ are normalizable, without caring about the
normalizability of the noncommutativity gauge fields $a$ and
$a\dag$, then far down the matrix $C$ and $\overline{C}$ will need
to converge to $a$ and $a\dag$ respectively. Therefore we can
treat $A$ and $\overline{A}$ as small perturbations and solve
(\ref{ddd}) order by order in $A$. The different orders in the
perturbation cannot mix far down the matrix, or else $A$ would
diverge. Notice that this approach differs from that of
Ref.~\cite{GN}, who did not impose (\ref{Cdef}) but rather imposed
the weaker condition that the covariant derivative satisfy a kind
of Leibniz rule and that the energy be finite. This led them to
extra superselection sectors of solutions, in which $C$ contains a
direct sum of $N$ copies of $a\dag$. These superselection sectors
were interpreted as $U(N)$ noncommutative gauge theories,
generalizing the $U(1)$ theory considered here.

Substituting (\ref{Cdef}) into the equation of motion (\ref{ddd})
we find \beq
0=[a+i\overline{A},[a\dag-iA,a+i\overline{A}]]=[a+i\overline{A},-1-\eta+[A,\overline{A}]]
\label{ddd2} \eeq where we have defined the Hermitian operator
\beq \eta=i[\overline{A},a\dag]-i[a,A]. \eeq Now we may expand
Eq.~(\ref{ddd2}) in powers of $A$ and take the linear term \beq
0=i[\overline{A},-1]-[a,\eta]=-[a,\eta] \eeq which implies that
$\eta$ is a function of $a$ \beq \eta=f(a). \eeq However $\eta$ is
Hermitian which implies that \beq f(a)=\eta=\eta\dag=\bar f(a\dag)
\eeq and so $f$ is a constant $c$ times the identity matrix.
Moreover $\eta$ is proportional to $A$, which goes to zero far
down the matrix, and so the constant of proportionality $c$ must
be zero, yielding \beq 0=\eta=i[\overline{A},a\dag]-i[a,A].
\label{notheta} \eeq

We will now restrict our attention to $\overline{A}$ of the form
\beq \overline{A}=[Q,a] \label{xformazione} \eeq and try to solve
for $Q$. This will give us a complete list of the continuous
symmetries of the solution that act via the adjoint representation
of a Lie group generated by the $Q$.
Substituting Eq.~(\ref{xformazione}) into (\ref{notheta}) yields
\beq
0=[[iQ,a],a\dag]+[a,[iQ\dag,a\dag]]=-i[[a,a\dag],Q]-i[[a\dag,Q],a]-i[a,[a\dag,Q\dag]]
\eeq where the first term on the right hand side vanishes because
$[a,a\dag]$ is proportional to the identity. This leaves \beq
[a,[a\dag,Q]]=[a,[a\dag,Q\dag]] \eeq and, subtracting the right
hand side from the left, one finds \beq [a,[a\dag,Q-Q\dag]]=0.
\eeq As a result $[a\dag,Q-Q\dag]$ is a function $f$ of $a$. The
commutator with $a\dag$ may also be written as the derivative with
respect to $a$, and so \beq \psi(a)=\frac{\partial}{\partial
a}(Q-Q\dag). \eeq The most general solution to this is \beq
Q-Q\dag=\lambda(a)+\rho(a\dag) \eeq where $\lambda$ and $\rho$ are
two functions, which establishes \beq Q=H+g(a)+h(a\dag)
\label{bog} \eeq where $H$ is Hermitian, and $g$ and $h$ are
arbitrary functions.

We may now interpret each term in $Q$ physically as a deformation
of the solution $\delta=a$. The Hermitian terms correspond to
ordinary $U(\ch)$ gauge transformations. If $g$ and $h$ are order
one polynomials then they describe a translation of the system.
Order two terms in $g$ and $h$ are Bogoliubov transformations.
Higher degree polynomials are even less normalizable than the
Bogoliubov transform.

\section{Adding commutative dimensions}

In the last section we searched for solutions to a 2-dimensional
noncommutative $U(1)$ gauge theory, which is equivalent to a
0-dimensional infinite-dimensional $PU(\ch)$ matrix model. We
found all solutions in which the adjoint scalar $\phi$ is a finite
codimension projector and the gauge field can be written as a
commutator of something with an annihilation operator. We found
the known solutions, their translations plus a series of
deformations of these solutions by nonnormalizable symmetries that
generalize Bogoliubov transformations. We identified these
solutions with stacks of $N$ 0-dimensional branes in a
2-dimensional background. We reproduced the fact that the blocks
of each component of the connection which is in the kernel of
$\phi$ are simultaneously diagonalizable and their eigenvalues
$\alpha$ and $\overline{\alpha}$ are arbitrary.

In the remainder of this paper we will be interested in a
generalization of this system which includes $d$ commutative
directions. In this new setting the complex combinations of the
gauge field in the two noncommutative directions $C$ and $\oC$ are
sometimes not diagonalizable when multiple branes are coincident.
We will see that in the case of charge 2 vortices the
nondiagonalizability is characterized by a single gauge-invariant
nonnegative real number, which corresponds to the radius of a
puffed vortex.

\subsection{Action and equations of motion}

We will be interested in the commutative $d$-dimensional $U(\ch)$
gauge theory which is equivalent to a $(d+2)$-dimensional $U(1)$
gauge theory on $R^{d+2}$ with two noncommuting directions and
adjoint matter. The $U(1)$ field strength has several new
nontrivial components, in addition to the old magnetic component
of Eq.~(\ref{mag}). If we use Greek indices to denote the
commutative directions $z^\mu$ and $w$ and $\overline{w}$ for the
noncommutative directions, then the mixed components of the field
strength are \beq F_{\mu w}=\partial_\mu A-\partial_w
A_\mu+i[A_\mu,A]=i\partial_\mu C+[A_\mu,-a\dag+iA]=i\partial_\mu
C-[A_\mu,C]=iD_\mu C \label{funa1} \eeq and similarly \beq
F_{\mu\overline{w}}=-iD_\mu\overline{C}. \label{funa2} \eeq
Letting uppercase Roman letters run over $z^\mu$,\ $w$ and
$\overline{w}$, the $(m+2)$-dimensional $U(1)$ gauge theory action
can be written (using the mostly minus metric) \beq
\label{azionegrande} S=\int dz^m d\overline{w}dw
(-\frac{1}{4}F_{MN}F^{MN}+\frac{1}{2}D_M\phi D^M\phi -V(\phi))
\eeq where the covariant derivative of $\phi$ is defined by \beq
D_M\phi=\partial_M\phi+i[A_M,\phi]. \eeq

We now expand the $w$ and $\overline{w}$ dependence of the fields
in a 2-dimensional basis of functions whose coefficients are
defined to be the fields in the $d$-dimensional $U(\ch)$ gauge
theory. We may express components of $F$ with one leg along a
noncommutative direction using (\ref{funa1}) and (\ref{funa2}) and
the component with both noncommutative legs using Eq.~(\ref{mag}).
Then the action (\ref{azionegrande}) can be written entirely in
terms of the infinite-dimensional matrices of the $d$-dimensional
theory
\begin{eqnarray}
S&=&\int dz^m
\Tr_{\ch}(-\frac{1}{4}F_{\mu\nu}F^{\mu\nu}+D_\mu\overline{C}D^\mu
C-\frac{1}{2}([C,\overline{C}]+1)^2\nonumber\\&&\ \ \ \ \ \ \ \ \
\ \ \ \ \ \ +\frac{1}{2}D_\mu\phi
D^\mu\phi-[C,\phi][\overline{C},\phi]-V(\phi)).
\end{eqnarray}

A complete set of equations of motion can now be found by setting
the variations with respect to $\phi$, $C$, $\overline{C}$ and
$A_\mu$ to zero. These variations respectively lead to the
following equations of motion
\begin{eqnarray}
D_\mu D^\mu\phi+[C,[\phi,\overline{C}]]+[\overline{C},[\phi,C]]+V\p(\phi)&=&0 \label{feq2}\\
D_\mu D^\mu\overline{C}+[\oC,[C,\oC]]+[\phi,[\oC,\phi]]&=&0 \label{ceq2}\\
D_\mu D^\mu C+[C,[\oC,C]]+[\phi,[C,\phi]]&=&0 \label{cbeq2}\\
D_\mu F^{\mu\nu}-i([C,D^\nu\oC]+[\oC,D^\nu
C]+[\phi,D^\nu\phi])&=&0 \label{aeq2}
\end{eqnarray}
which reduce to Eqs.~(\ref{feq},\ref{ceq},\ref{cbeq}) in the case
$d=0$ as they must.

The solutions of the $d=0$ matrix theory case easily generalize to
solutions in this case, one need only assert that all fields are
constant along the $d$ commutative directions $z^\mu$ and that
$A_\mu$ is identically zero. These generalizations correspond to
stacks of $N$ flat codimension two branes. When the functions $g$
and $h$ in the solution (\ref{bog}) are zero, then these branes
are centered at the origin of the $w$, $\overline{w}$ plane. Not
all such configurations are gauge equivalent, because one must
still choose the $N$ complex eigenvalues $\alpha_i$ which yields a
moduli space $\C^N/\Z_N$ of seemingly inequivalent brane
configurations, these configurations are related by a global
symmetry, although they are related by a gauge symmetry up to an
arbitrarily small correction. It is tempting to identify this
space of Wilson lines with the positions of the branes, however
such positions are in general not well-defined in a noncommutative
gauge theory and so instead in the next section we will use these
eigenvalues as definitions of the positions. We will see below
that if one relaxes the $z$-independence of $C$ then the
$\alpha_i$ lead to a kind of electric dipole moment despite the
lack of electric charges in the solution and even to puffed
solutions of $N$ vortices in which the upper-left $N\times N$
block of $C$ is not diagonalizable.

\subsection{Electric dipoles and polarized branes}

Consider the aforementioned $z$-independent solution with
$g=h=A_\mu=0$, describing $N$ straight codimension 2 branes
extending along the $z$ directions with a trivial longitudinal
connection. We have noted that these configurations are
parameterized by the complex eigenvalues $\alpha_i$. Now allow the
$\alpha_i$ to depend on $z$. In Sec.~\ref{puffi} we will consider
solutions in which the $\alpha$ block is not diagonalizable, for
now we will restrict our attention to solutions in which it is,
and we will consider a basis in which it is diagonal and we will
furthermore set all commutative components of the gauge field to
zero, as well as off-block diagonal components of the gauge field
in the noncommutative directions, which correspond to tachyonic
instabilities \cite{GN}. This leaves us with the solutions \beq
\label{indep} \phi=\phi_*(1-P_N)\hsp C(z)=\sum_{i=0}^{N-1}
\alpha_i(z)|i\rangle\langle i|+ S^Na^\dagger\bar S^N\hsp
A_\mu(z)=0. \eeq

Now the $\alpha_i$ appear on the diagonal of $C$ and so commute
with $\phi$, $C$ and $\overline{C}$. Therefore they do not
contribute to Eq.~(\ref{feq2}) and they only contribute to the
first term in Eqs.~(\ref{ceq2},\ref{cbeq2}). As $A_\mu=0$, the
$\alpha_i$ also do not contribute to Eq.~(\ref{aeq2}). Thus the
only constraint on the $\alpha_i$ comes from the first term of
Eqs.~(\ref{ceq2},\ref{cbeq2}), which in the case $A_\mu=0$ reduces
to the wave equation \beq
\partial_\mu\partial^\mu \alpha_i=0 \label{vague}
\eeq as noted in Ref.~\cite{DN}.

To interpret the solutions, first consider the special case $d=1$.
As the signature of the spacetime does not affect the formal
considerations here, we will consider the single commutative
direction to be the time $t$. The wave equation (\ref{vague}) then
implies that the $\alpha_i$ are linear \beq \alpha_i=c_i+d_it.
\eeq The $c_i$ are Wilson lines as in the time-independent case.
The new elements are the $d_i$. As the $\alpha_i$ are diagonal,
they resemble the vector potentials for the $U(1)^N\subset U(N)$
gauge group that lives on the stack of $N$ branes, except that
they are perpendicular to the worldvolumes.

The electric fields on the worldvolumes are the time derivatives
of the vector potentials, therefore the $i$th brane has an
electric field \beq E_x=d_i+d_i^*\hsp E_y=i(d_i-d_i^*). \eeq These
electric fields are not parallel to the branes, they are
orthogonal, and so they are also not a part of the worldvolume
gauge theory. They are instead worldvolume electric fields in the
$U(1)$ gauge theory of the spacefilling brane. In the worldvolume
theory of the spacefilling brane, the codimension 2 branes are
magnetic vortices. The $d_i$ imply that in addition to a magnetic
flux running along the brane, there is also an electric flux
perpendicular to the brane. In other words, the branes have an
electric dipole moment, despite the fact that there is no
electrically charged matter in the theory except for the
off-diagonal components of the gluons, whose values are equal to
zero in this solution. This appears to be a novel phenomenon in
noncommutative gauge theories, an electric dipole moment can exist
without a source. In the commutative limit it smears out and
becomes a constant electric flux which is supported by boundary
conditions, but in the noncommutative case it exists as a
localized lump.

The field strength of a magnetic flux tube in a commutative gauge
theory is perpendicular to the tube. The $d_i$ component on the
other hand has one leg perpendicular to the tube and one leg along
the tube. Thus the total field strength 2-form is slanted, along
an axis determined by the phase of $d_i$ and by an amount
proportional to the arctangent of the magnitude of $d_i$. We will
refer to such solutions as {\it polarized branes}.

Returning to the case of an arbitrary number of commutative
directions $d$, the derivative of $\alpha_i$ in each commutative
direction is a magnetic flux component perpendicular to the
magnetic vortex, therefore again we find polarized branes. However
when $m>1$, Eq.~(\ref{vague}) admits wave solutions, and so the
perpendicular polarizations of the magnetic fields propagate.



\subsection{What is position?}

A noncommutative spacetime is not really composed of points, in
the sense that there are gauge transformations which translate any
field which transforms in a nontrivial representation of a gauge
group. Technically translations of the whole spacetime are not
gauge symmetries because they do not vanish at infinity
\cite{Harveyuh} and so can be fixed by the boundary conditions of
the path integral. However we will be interested in vortex
solutions which, at least almost everywhere, vanish at infinity in
the noncommutative directions and such solutions may be translated
by legitimate gauge transformations which fall off sufficiently
quickly at infinity.

Our solutions are composed of two fields, the gauge field and the
adjoint scalar, both of which transform in the adjoint
representation of the gauge group. Explicitly, via a global
transformation
\begin{eqnarray} \label{globale}
\phi&\longrightarrow& e^{wa\dag-w\dag a}\phi e^{-wa\dag+w\dag a}\\
C&\longrightarrow& e^{wa\dag-w\dag a}C e^{-wa\dag+w\dag a}\nonumber\\
A_\mu &\longrightarrow& e^{wa\dag-w\dag a}(A_\mu-i\partial_\mu)
e^{-wa\dag+w\dag a}\nonumber
\end{eqnarray}
one can move the core of a vortex to any codimension 2 submanifold
of spacetime that intersects each noncommutative plane precisely
once without changing the energy of the solution. This is in
contrast with the commutative case in which one expects that the
energy depends on the volume of the vortex. Even more seriously,
one may truncate this global transformation by projecting it with
a projector whose rank is much higher than the charge of the
vortex, in this case the truncated action on the vortex will be
arbitrarily close to that of the global transformation
(\ref{globale}), but it will be a gauge transformation. Therefore
vortices whose $\phi, C$ and $A_\mu$ profiles have dramatically
different centers, for example straight branes and sine curves,
are gauge-equivalent.

Therefore it appears that the position of a vortex in a
noncommutative gauge theory is a gauge artifact. However, in
Ref.~\cite{GN} Gross and Nekrasov point out that gauge-invariant
operators have well-defined distributions. Therefore they define a
gauge-invariant notion of the spatial distribution of a soliton as
the distribution of these gauge-invariant operators. They quickly
ran into the problem that the different gauge-invariant operators
that they defined did not agree on the form-factors of the
internal structure of the vortex, however they did provide an
apparently well-defined notion of the location of the core of the
vortex, at least in the case in which the vortex's position is
independent of time and space. Recall that in this case the
top-left $N\times N$ block of the $C$ matrix of a charge $N$
vortex is diagonalizable with eigenvalues $\alpha_i$. They found
that their gauge-invariant operators are centered on $N$ points on
the complex $w$ plane, which are equal to the $N$ eigenvalues
$\alpha_i$, as had already been conjectured in Ref.~\cite{AGMS}
based on an analogy with matrix theory. Subsequent authors
\cite{DN,Szabo,Szabo2} adopted the claim of \cite{AGMS} that this
result extends to solutions which are not uniform in the
commutative directions. The identification of position with Wilson
lines resembles the T-dual position of a D-brane that wraps a
circle, however in this case the vortex does not actually extend
along the noncommutative directions.

We will now momentarily restrict our attention to the class of
solutions (\ref{indep}). The commutative functions $\alpha_i(z)$
are solutions to the wave equations (\ref{vague}). Hence they can
be interpreted as minimal area codimension 2 worldvolumes for
lower dimensional D-branes. While in the pure noncommutative
theory the $\alpha_i$'s are actually moduli of the solution, here
they are solutions to the d'Alembert equations and so the
time-independent solutions are characterized by the momenta of
their Fourier transforms. The functions $\alpha_i$ are eigenvalues
of the tensor $C$ and so are gauge invariant, therefore they
define in an unambiguous way the actual positions of the lower
dimensional D-branes in the transverse direction. We note however
that these solutions solve the equations of motion in a somewhat
trivial way. Indeed every monomial in the equations of motion
vanishes individually.

Gross and Nekrasov defined the positions of these vortices using
two distinct gauge-invariant operators. First, they considered
adding fundamental fermion probes $\psi$ to the theory. While the
fermion field $\psi$ itself is gauge-dependent, and so its
position is ill-defined, the bilinear $\psi\dag\psi$ is a gauge
singlet. Fermions satisfy the Dirac equation, which in the
noncommutative dimensions is
\begin{eqnarray}
0&=&\tilde D\psi_1=\partial_{\overline{w}}\psi_1+i\oA\psi_1=[w,\psi_1]+i\oA\psi_1=\oC\psi_1-\psi_1 w\nonumber\\
0&=&D\psi_2=\partial_w\psi_2+iA\psi_2=-[\overline{w},\psi_2]+iA\psi_2=-C\psi_2+\psi_2
\overline{w} \label{dirac}
\end{eqnarray}
where $\psi_1$ and $\psi_2$ are left and right handed Weyl
fermions and $a=w,\,a^\dagger=\bar w$ . Notice that although the
fermions transform in the fundamental representation of the gauge
group, the Weyl transformation means that they are represented by
matrices and transform in the adjoint of the Heisenberg algebra.
Using this definition of fundamental fermions, in which the
connection acts on these matrices on the left, the covariant
derivatives of the left handed and right handed fermions are not
conjugates. This would have been the case if instead one had
imposed that $\oA$ act on $\psi_1$ via right multiplication.
Perhaps such a definition of fundamental fermions would be
interesting to investigate.

The normalizable zeromodes consist of matrices $\psi_1$ whose
right eigenvalues under the position operator $\overline{w}$ are
equal to the eigenvalues $\alpha_i$ of the connection $C$.
Therefore the eigenvalues of the position operator $\overline{w}$
on the bilinear $\psi\dag\psi$ from either the right or left are
just the $\alpha_i$, and so a charge $N$ vortex has $N$ fermion
zero modes whose wavefunctions are each centered at the position
corresponding to the eigenvalue $\alpha_i$. Intuitively, the Dirac
equation (\ref{dirac}) just imposes that the position $w$ of a
fermion charged under a particular $U(1)$ is just equal to the
Wilson line $\overline{\alpha_i}$ of that $U(1)$. As the fermion
position is gauge invariant, Gross and Nekrasov then define the
location of the fermions to be the location of the vortex. They
also define the density of the vortex to be that of the fermions,
which they found to be Gaussians of width $\sqrt{\theta}$.

In the case of puffed branes we will see that some of the
fermionic zeromodes are lifted, and the others are invariant under
the puffing parameter. Instead the gauge-invariant data will be
captured by another gauge-invariant operator, which in the case of
the solutions (\ref{indep}) is also centered on the points
$\alpha_i$, although it is focused at delta functions and in fact
its normalization is not canonically defined and so with a
suitable choice of normalization it can yield any form factors for
the brane.

The trace of $C$ is gauge-invariant, and it captures the center of
mass of the vortices. To capture the positions of all of the
vortices, \cite{GN} consider instead a kind of Wilson loop
\be\label{wilson1} W(q)=\tr\left(e^{\bar q C- q \bar C}\right).
\ee Note that even if the $C$ field is not Hermitian, the exponent
of the Wilson loop is anti-Hermitian. Ordinarily a Wilson loop is
integrated over a closed loop. In noncommutative space the trace
of an infinite-dimensional matrix $C$ is the same as the integral
of the Weyl transformed $U(1)$ gauge field. Therefore $W(q)$ is a
kind of integral over all of the Wilson lines oriented in the $q$
direction, weighted by $|q|$. The definition of the distribution
of a vortex in Ref.~\cite{GN} is that the momentum space
distribution be identified with $W(q)$. They then identify the
Fourier transform of $W(q)$ with the position distribution, which
is the Fourier transform of a plane wave. A quick calculation
shows that this is just a sum of Dirac delta functions centered at
the points $\alpha_i$ on the noncommutative plane.

\section{Puffed vortices} \label{puffi}

In Ref.~\cite{GHS} the authors find that the moduli space of
vortices in a noncommutative scalar field theory is modified by
$\theta$ corrections when two vortices intersect. As a result the
expected singularity is blown up into a compact, nonsingular
projective space. We consider a noncommutative gauge theory to all
orders in $\theta$ and find, not surprisingly, a different moduli
space of solutions. Again we find that the moduli spaces of
solutions has an extra degree of freedom when multiple vortices
coincide, however in the present case we will see that this extra
direction is noncompact as in the case of semilocal vortices
\cite{semilocal,semilocal2}. In the case of charge 2 vortices
there is a single additional nonnegative real gauge-invariant
quantity. Defining the vortex profile using the Wilson lines
introduced by Gross and Nekrasov, we will see that coincident
vortices puff up into rings of domain walls as in
Refs.~\cite{Stefano,Stefano2,Stefano3} and that the extra
parameter is the radius of the ring.

To find the puffed vortex solutions, we will slightly relax our
ansatz (\ref{indep}) by allowing the $N\times N$ upper left block
of $C$ to be arbitrary, without changing the other components and
without changing $\phi$. In this case there are gauge-inequivalent
new solutions only when the minimal polynomial of its Chan-Paton
matrix has degenerate linear factors, corresponding intuitively to
degenerate eigenvalues, otherwise one can always bring it back in
diagonal form by using a $U(N)\subset U(\ch)$ gauge
transformation. We will now restrict our attention to charge 2
vortices, so that $C$ has a 2 dimensional degenerate eigenspace.
Unlike $\phi$, $C$ does not have to be Hermitian and if it is not
diagonalizable then it is not Hermitian, however a $U(2)$ gauge
transformation can bring it into the Jordan form
\beq\label{newansatz} C_2(z)=\left(
\begin{array}{cc}
\alpha(z)&\beta(z)\\
0&\alpha(z)
\end{array}\right).
\eeq

By performing a gauge transformation generated by $\sigma_3$ it is
easy to see that the phase of $\beta(z)$ is a gauge artifact. Only
the modulus $|\beta|$ is a gauge-invariant quantity. For
simplicity we will partially fix the gauge by choosing the Jordan
canonical form of the $2\times 2$ upper-left block of $C$ with
$\beta$ real, but we will not use the gauge freedom to force
$\beta$ to be positive, as the corresponding gauge transformations
will not be differentiable in the space-dependent solutions below.

\begin{figure} \label{wilfig}
\begin{center}
\includegraphics[width=10cm,angle=0] {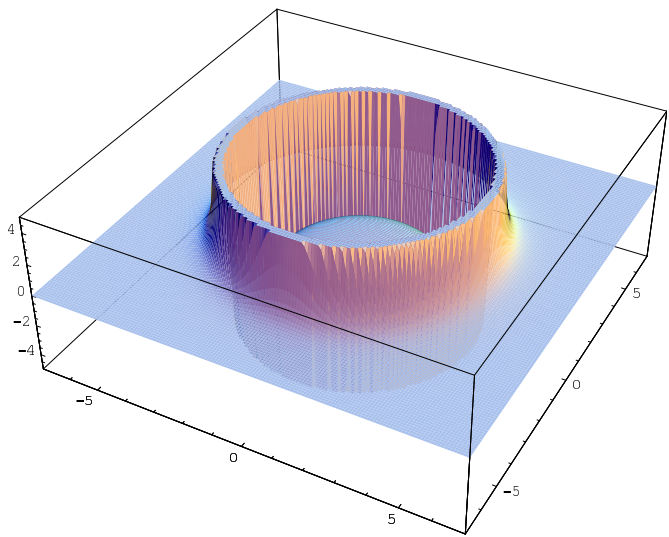}
\caption{Profile of the Wilson lines in the presence of a puffed
brane. The Wilson lines are largest on a ring of radius equal to
the gauge-invariant parameter $\beta$.} \label{whatever name-tag
you want to give the figure if you want to refer to it in the
article}
\end{center}
\end{figure}

We will now use the Wilson loops of Ref.~\cite{GN} to attempt to
physically interpret the gauge-invariant parameter $|\beta|$.
From now on we will just concentrate on the rank 2 block
containing $\beta$. The Wilson loop is \be e^{\bar q C- q \bar
C}=e^{\bar q \alpha- q \bar \alpha}\,\cos(|q|\,\beta)\left(
\begin{array}{cc}
1&\tan(|q|\,\beta)\frac{\bar q}{|q|}\\
-\tan(|q|\,\beta)\frac{q}{|q|}&1
\end{array}\right) .
\ee Taking the trace, the (additive) contribution from this block
is \be w(q)=2\, e^{\bar q \alpha- q \bar
\alpha}\,\cos(|q|\,\beta)\0. \ee According to the Gross-Nekrasov
prescription, $q$ should now be identified as a gauge-invariant
momentum coordinate for the state described by the Wilson loop. To
relate $\beta$ to the position of the vortex, we must first
Fourier transform $w(q)$ to position space \be \hat
w(x)=\frac1{(2\pi)^2}\int\, d^2 q\,w(q)\,e^{i(q\bar x + \bar q
x)}=\frac2{(2\pi)^2}\int\,d^2q\,e^{i(\bar q x_\alpha+q\bar
x_\alpha)} \cos(|q| \beta) \ee where we have defined the complex
position $x_\alpha=x-i\alpha$.

Going to polar coordinates $q=\rho\, e^{i\theta}$,
$x_\alpha=r\,e^{i\phi}$, splitting the cosine in two exponentials,
shifting the integration variable $\theta\to\theta+\phi$ and
introducing an $\e$-cutoff for the radial integration we get \be
\hat w(x)=\frac1{(2\pi)^2}\int d\theta\,\rho\,d\rho
\left(e^{[i(r\,\cos\theta+\beta)-\e]\rho}+e^{[i(r\,\cos\theta-\beta)-\e]\rho}\right)
\ee Keeping $\epsilon$ only where it is needed, the integral can
be evaluated analytically \be \hat
w(x)=-\frac1{2\pi}\,\frac\beta{(\beta+r)^{\frac32}}\left(\frac1{(\beta-r-i\e)^{\frac32}}+\frac1{(\beta-r+i\e)^{\frac32}}\right).
\ee

Note that this field configuration is real, as it should be, and
it is axially-symmetric with respect to the position of the
$\beta$-unperturbed vortices $x=i\alpha$. The distribution $\hat
w(x)$ describes a thin circular shell of radius $\beta$
surrounding the original vortices positions, as seen in Fig.~1.
Turning $\beta$ down to zero the shell shrinks and one arrives at
the old solution describing two coincident pointlike vortices at
$i\alpha$.

Recall that there is a second gauge-invariant definition of the
spatial profile of a solution, one may couple the system to
fundamental fermion probes and use the profile of the fermion
bilinears. If one includes a single additional dimension $z$, then
one may probe the system with a 3-dimensional Dirac fermion $\Psi$
with two complex (matrix valued) components $\psi_1$ and $\psi_2$
which satisfy the Dirac equation \be
\partial_z\psi_1-C\psi_2+\psi_2 a\dag&=&0\0\\
-\partial_z\psi_2+\oC\psi_1-\psi_1a&=&0. \ee In general a rank $N$
vortex solution has $N$ fermion zero modes if it is independent of
the $z$ coordinate. However, as we will explain presently, when
$\beta$ is nonzero the equations of motion demand that the
vortices always depend on the $z$ coordinate. This lifts some of
the fermion zero modes.

For example, in the case of the puffed charged two vortices that
we have analyzed, when $\beta$ is nonzero one of the fermion zero
modes is unaffected, the generator of the zero-eigenspace of $C$,
while the other if lifted. The unaffected fermion is insensitive
to $\beta$, and so cannot be used as a probe, it is always equal
to the tensor product of the oscillator ground state with a
coherent state with coefficient equal to the position of the
center of the puffed vortex, so that its bilinear is an eigenstate
of the position operator centered in the center of the vortex.

The fate of the lifted zeromode is more interesting. While it is
no longer a zeromode, one may find an exact solution to the Dirac
equation that describes its evolution in the commutative
direction. If, for simplicity, the puffed vortex is centered on
the origin, then the individual matrix elements of the fermion are
not all coupled in the Dirac equation, they appear in isolated
groups of 4, corresponding to various background configurations.

The lifted zeromode only appears in one of these groups. If we
write explicitly the matrix form of the components of the Dirac
fermion $\Psi$ as \be
\psi_{1,2}=\sum_{q=0}^\infty\left(\begin{array}{c}\rho_{1,2}^q |0\ket \bra q|\\
\gamma_{1,2}^q |1\ket\bra q|\end{array}\right) \ee then the Dirac
equation for the lifted zeromode is simply \be
\partial_z \rho^0_1&=&\beta\,\gamma^0_2\0\\
\partial_z\gamma_1^1&=&\gamma_2^0\0\\
\partial_z\gamma_2^0&=&\beta\,\rho_1^0-\gamma_1^1.
\ee This system of homogeneous linear differential equations is
solved by linear combinations of three generalized hypergeometric
functions. However no linear combination of these functions
appears to be normalizable for the solutions found below, and so
we cannot use the locations of the fermion bilinears to define the
positions of our solutions, as was possible in the $\beta=0$ case
studied by Gross and Nekrasov.

Now that we have understood the meaning of $\beta$, we may attack
the problem of its evolution under the equations of motion.
Inserting the ansatz (\ref{newansatz}) into the equations of
motion, one sees that the $\phi$ equation is unaffected, and the
same is true for the $A_\mu$'s (if $\beta$ is not real, then one
should compensate for its phase with a corresponding pure gauge
shift in $A_\mu$). The only nontrivial equation is that obtained
by varying $C$. Assuming that for $\beta=0$ we have a solution
(that is $\Box \alpha=0$), turning on $\beta$ implies the equation
\beq \Box \beta +2\beta^3=0. \eeq We do not know how to solve this
equation in general, but we will study two treatable cases.
\begin{figure}
\begin{center}
\includegraphics[width=6.5cm,angle=0] {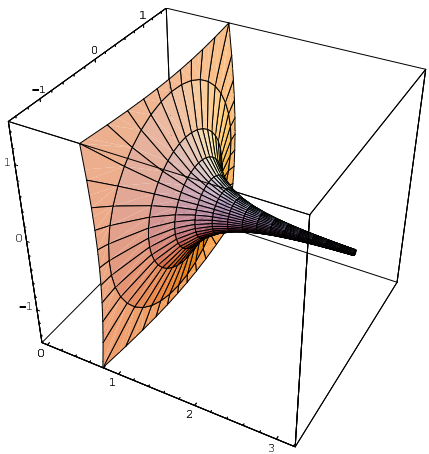}
\caption{Time-independent solutions in which the radius
$\beta$ depends on a single spatial dimension $x$: they are funnels in which
the radius $\beta$ reaches infinity at a finite $x$. These
solutions are smooth and describe a vortex ending on a domain
wall. These are limits of solutions describing a vortex stretched between two parallel domain walls} \label{tempofig}
\end{center}
\end{figure}

First we classify static solutions in which $\beta$ depends only
on a single spatial variable $z^i=x$. The equation then becomes
\beq
\frac{d^2\beta}{dx^2}=2\beta^3
\eeq
Bearing in mind that only
the modulus of $\beta$ is gauge-invariant, a one parameter family
of solutions is given by
\beq\label{wall}
|\beta(x)|=\frac{1}{|x-x_0|}\ .
\eeq

This solution has two distinct branches, one at $x<x_0$ and one at
$x>x_0$, which is depicted in Fig.~2. Each branch describes a
D23-brane that ends on a D24.

Another interesting class of solutions arises when we take $\beta$
to be time-dependent ($z^0=t$) but spatially homogeneous. In this
case the differential equation becomes
\beq
\frac{d^2\beta}{dt^2}=-2\beta^3. \label{espace}
\eeq
A solution to this equation is given by
\beq
\beta(t)=a\,{\textrm{dn}}\!\left(a(t-t_0),\,2\right)
\eeq
where dn is a Jacobi elliptic function.

This solution looks roughly like a string of ellipsoids
attached end to end, although the derivative of $\beta$ is always
finite so each intersection is just the union of two opposing
cones that touch at their tips, as illustrated in Fig.~3. Each
ellipsoid represents a bubble nucleating at some time $t_0$,
expanding to some maximal size and then decaying back to a point
after a period.
Note that the Wick rotation of such solutions,
\be\label{array}
\beta(x)=b\,{\textrm{dn}}\!\left(ib(x-x_0),\,2\right),
\ee
gives a real space--dependent solution, which have periodic singularities on an array defined
by the initial conditions, the period being $2/b K(-1)$, with $K(q)$ the complete elliptic integral of the first kind.
Such extra solutions can be interpreted as D23 branes stretched between two D24's.
On the other hand the special space-solution (\ref{wall}), does not admit a real inverse Wick rotation, so it does not generate an extra time
dependent solution. Notice moreover that this special solution is obtained from (\ref{array}) by taking $x_0=K(-1)/b$,
and sending $b\to0$.
\begin{figure}
\begin{center}
\includegraphics [width=10cm,angle=0] {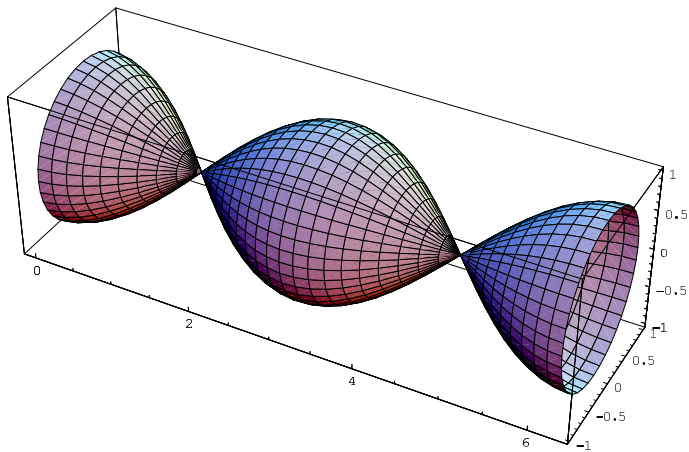}
\caption{Spatially-homogeneous time-dependent solution describing
a braneworld cyclic universe} \label{salsiccie}
\end{center}
\end{figure}

Such solutions generally have tachyonic instabilities at the tips
of the cones, but these are of little concern here as the vortices
in this note all suffer from tachyonic instabilities everywhere.
They can be stabilized, for example, if one modifies the potential
energy so as to spontaneously break the $U(1)$ gauge symmetry at
infinity.



Notice that the adjoint scalar plays no role in these vortex
solutions, although their forms are fixed by the charge of the
vortex. The solutions discussed in this note are therefore also
solutions to the pure gauge theory. We expect that many other
interesting solutions can be found.

As a side remark we would like to comment on the relation with
Open String Field Theory, observing that working at finite
$\theta$ seems to mark a profound difference with respect to the
infinite noncommutative limit. In particular, at infinite $\theta$
there is a one to one correspondence between noncommutative
solitons and String Field Theory solutions \cite{BMS, carlo,
martin} (where the string field is basically the noncommutative
tachyon). At finite $\theta$, on the other hand, there is no way
to get rid of the gauge connection, and the tachyon field plays a
minor role (and could be even thrown away without changing the
relevant physics). From a formal field theory point of view this
is for us no surprise as both the gauge field and the open string
field are the connection of two infinite dimensional gauge groups
that share lot of similarities \cite{Harveyuh}. While it is
understood that $U(1)$ noncommutative gauge theory is the low
energy limit of OSFT on a D25--brane with a constant $B$-field on
its worldvolume, it seems that the two theories are very different
in the way they are classically solved. It would be therefore
interesting to understand the relations (if any) between classical
solutions of the two theories, in particular to find the OSFT
counterpart of our puffed solutions. It is clear that this
question will only be addressable once classical solutions for multiple
and lower dimensional D-branes will be understood in OSFT.

\vskip 1cm

\begin{center}
{\bf Acknowledgments}
\end{center}
We would like to express our gratitude to L. Bonora and R. Szabo for carefully reading our manuscript.
We also thank  G. Barnich, F. Ferrari, C. Krishnan, S. Kuperstein and  even D. Persson for enlightening comments and discussions.

N.B. and J.E. are supported in part by a ``Pole d'Attraction Interuniversitaire''
(Belgium), by IISN-Belgium, convention 4.4505.86, by Proyectos
FONDECYT 1970151 and 7960001 (Chile) and by the European
Commission program
MRTN-CT-2004-005104, in which this author is associated to V.U. Brussels.\\
C.M. is supported in part by the Belgian Federal Science Policy
Office through the Interuniversity Attraction Pole P5/27, in part
by the European Commission FP6 RTN programme MRTN-CT-2004-005104
and in part by the
``FWO-Vlaanderen'' through project G.0428.06.\\


\end{document}